\documentstyle[preprint,aps,epsf]{revtex}

\def\pmb#1{\setbox0=\hbox{$#1$}%
\kern-.025em\copy0\kern-\wd0
\kern.05em\copy0\kern-\wd0
\kern-.025em\raise.0433em\box0}

\def\beq{\begin{equation}}
\def\eeq{\end{equation}}

\begin{document}

\def\footnoterule{\hrule width \hsize}
\def\footstrut{\baselineskip 16pt}

\skip\footins = 14pt 
\footskip     = 20pt 
\footnotesep  = 12pt 

\textwidth=6.5in
\hsize=6.5in
\oddsidemargin=0in
\evensidemargin=0in
\hoffset=0in

\textheight=9.5in
\vsize=9.5in
\topmargin=-.5in
\voffset=-.3in

\baselineskip=24pt plus .5pt

\title{%
THE UNREASONABLE EFFECTIVENESS OF \\ 
QUANTUM FIELD THEORY%
}

\footnotetext[1] {\baselineskip=16pt This work is supported in part by funds
provided by  the U.S.~Department of Energy (D.O.E.) under contract
\#DE-FC02-94ER40818. \hfil MIT-CTP-2500 \hfil  January 1996\break}

\author{R.~Jackiw\footnotemark[1]}

\address{Center for Theoretical Physics\\ Massachusetts Institute of
Technology\\ Cambridge, MA ~02139--4307}

\maketitle

\setcounter{page}{0}
\thispagestyle{empty}



\vskip 5in

\centerline{``70 Years of Quantum Mechanics,'' Calcutta, India, January 1996}

\centerline{``Foundations of Quantum Field Theory,'' Boston, MA, March 1996}

\maketitle

\newpage

Quantum field theory offers physicists a tremendously wide range of
application; it is both a language with which a vast variety of physical
processes can be discussed and also it provides a model for fundamental
physics, the so-called ``standard-model,'' which thus far has passed every
experimental test.  No other framework exists in which one can calculate so
many phenomena with such ease and accuracy.  Nevertheless, today some
physicists have doubts about quantum field theory, and here I
want to examine these reservations.  So let me first review the successes.

Field theory has been applied over a remarkably broad energy range
and whenever detailed calculations are feasible and justified, numerical
agreement with experiment extends to many significant figures.  Arising from a
mathematical account of the propagation of fluids (both ``ponderable'' and
``imponderable''), field theory emerged over a hundred years ago in the
description within classical physics of electromagnetism and gravity
[\ref{ref:1}].  Thus its first use was at macroscopic energies and distances,
with notable successes in explaining pre-existing data (relationship between
electricity and magnetism, planetary perihelion precession) and predicting new
effects (electromagnetic waves, gravitational bending of light). 
Schr\"odinger's wave mechanics became a bridge between classical and quantum
field theory: the quantum mechanical wave function is also a local field,
which when ``second'' quantized gives rise to a true quantum field theory,
albeit a non-relativistic one.  This theory for atomic and chemical processes
works phenomenally well at electron-volt energy scales or at distances of
O(10$^{-5}$cm).  Its predictions, which do not include relativistic and
radiative effects, are completely verified by experiment.  For example the
ground state energy of helium is computed to seven significant figures; the
experimental value is determined with six figure accuracy; disagreement, which
is seen in the last place, disappears once radiative and relativistic
corrections are included.  Precisely, in order to incorporate relativity and
radiation, quantum field theory of electromagnetism was developed, and
successfully applied before World War II to absorption or emission of real
photons.  Calculation of virtual photon processes followed the war, after
renormalization theory succeeded in hiding the infinities that appear in the
formalism. [\ref{ref:2}]  Here accuracy of calculation is achieved at the
level of one part in 10$^{8}$ (as in the magnetic moment of the electron,
whose measured  value agrees completely with theory), where distances
of O(10$^{-13}$cm) are being probed.  Further development, culminating with the
standard particle physics model, followed after further infinities that
afflict theories with massive vector mesons were tamed.  Indeed the masses of
the vector mesons mediating weak interactions (by now unified with
electromagnetism) were accurately predicted, at a scale of 100 $GeV$.

I have summarized briefly quantum field theoretic successes within elementary
particle physics.  It is also necessary to mention the equally impressive
analyses in condensed matter physics, where many fascinating
effects (spontaneous symmetry breaking, both in the Goldstone-Nambu and
Anderson-Higgs modes, quantum solitons, fractional charge, {\it etc}.) are
described in a field theoretic language, which then also informs elementary
particle theory, providing crucial mechanisms used in the latter's model
building.  This exchange of ideas demonstrates vividly the vitality and
flexibility of field theory.  Finally we note that quantum field theory has
been extrapolated from its terrestrial origins to cosmic scales of distance
and energy, where it fuels ``inflation'' -- a speculative, but completely
physical analysis of early universe cosmology, which also appears to be
consistent with available data.

With this record of accomplishment, why are there doubts about quantum field
theory, and why is there vigorous movement to replace it with string theory,
to take the most recent instance of a proposed alternative?  Several reasons
are discerned.  Firstly, no model is complete -- for example the standard
particle physics model requires {\it ad hoc\/} inputs, and does not encompass
gravitational interactions.  Also, intermittently, calculational
difficulties are encountered, and this is discouraging.  But these
shortcomings would not undermine faith in the ultimate efficacy of quantum
field theory were it not for the weightiest obstacle: the occurrence of
divergences when the formalism is developed into a computation of physical
processes.

Quantum field theoretic divergences arise in several ways.  First of all,
there is the lack of convergence of the perturbation series, which at best is
an asymptotic series.  This phenomenon, already seen in quantum mechanical
examples like the anharmonic oscillator, is a shortcoming of an approximation
method and I shall not consider it further. 

More disturbing are the infinities that are present in every perturbative
term, beyond the first.  These divergences occur after
integrating or summing over intermediate states -- a necessary calculational
step in every non-trivial perturbative order.  When this
integration/summation is expressed in terms of an energy variable, an
infinity can arise either from the infrared  -- low energy -- and/or
from the ultraviolet -- high energy -- domains. 

 The former, infrared infinity
afflicts theories with massless fields and is a consequence of various
idealizations for the physical situation:  taking the region of space-time,
which one is studying, to be infinite and supposing that massless particles
can be detected with infinitely precise energy-momentum resolution are
physically unattainable goals and lead in consequent calculations to the
afore-mentioned infrared divergences.  In quantum electrodynamics one can
show that physically realizable experimental situations are described within
the theory by infrared-finite quantities.  Admittedly, thus far we have not
understood completely the infrared structure in the non-Abelian
generalization of quantum electrodynamics -- this generalization is an
ingredient of the standard model -- but we believe that no physical
instabilities lurk there either.  So the consensus is that infrared
divergences, do not arise from any intrinsic defect of the theory, but rather
from illegitimate attempts at forcing the theory to address unphysical
questions.

Finally, we must confront the high-energy, ultraviolet infinities.  These
{\bf do} appear to be intrinsic to quantum field theory, and no physical
consideration can circumvent them: unlike the infrared divergences,
ultraviolet ones cannot be excused away.  But they {\bf can} be
``renormalized.''  This procedure allows sidestepping or hiding the infinities
and succeeds in unambiguously extracting numerical predictions from the
standard model and from other ``physical'' quantum field theories, with the
exception of Einstein's gravity theory -- general relativity -- which thus far
remains ``non-renormalizable.''

The apparently necessary
presence of ultraviolet infinities has dismayed many who remain unimpressed
by the pragmatism of renormalization: Dirac and Schwinger, who count among
the creators of quantum field theory and renormalization theory,
respectively, ultimately rejected their constructs because of the infinities. 
But even among those who accept renormalization, there is disagreement about
its ultimate efficacy at well-defining a theory.  Some argue that sense can be
made only of ``asymptotically free'' renormalizable field theories (in these
theories the interaction strength decreases with increasing energy).  On the
contrary, it is claimed that asymptotically non-free models, like
electrodynamics and
$\phi^4$-theory, do not define quantum theories, even though they are
renormalizable -- it is said ``they do not exist.''  Yet electrodynamics
is the most precisely verified quantum field theory, while the $\phi^4$-model
is a necessary component of the standard model, which thus far has met no
experimental contradiction.

The ultraviolet infinities appear as a consequence of space-time localization
of interactions, which occur at a point, rather than spread over
a region.  (Sometimes it is claimed that field theoretic infinities arise
from the unhappy union of quantum theory with special relativity.  But this
does not describe all cases -- later I shall discuss a non-relativistic,
ultraviolet divergent and renormalizable field theory.)  Therefore choosing
models with non-local interactions provides a way for avoiding ultraviolet
infinities.  The first to take this route was Heisenberg, but his model was
not phenomenologically viable.  These days in string theory non-locality is
built-in at the start, so that all quantum effects -- including gravitational
ones -- are ultraviolet finite, but this has been achieved at the expense of
calculability: unlike ultravioletly divergent local quantum field theory,
finite string theory has not yielded even an approximate calculation of any
physical process.

My goal in this talk is to persuade you that the divergences of quantum field
theory must not be viewed as unmitigated defects; on the contrary, they convey
crucially important information about the physical situation, without which
most of our theories would not be physically acceptable.  The stage where such
considerations play a role is that of symmetry, symmetry breaking and
conserved quantum numbers, so next I have to explain these ideas.

Physicists are mostly agreed that ultimate laws of Nature enjoy a high degree
of symmetry.  By this I mean that the formulation of these laws, be it in
mathematical terms or perhaps in other accurate descriptions, is unchanged
when various transformations are performed.  Presence of symmetry implies
absence of complicated and irrelevant structure, and our conviction that this
is fundamentally true reflects an ancient aesthetic prejudice -- physicists
are happy in the belief that Nature in its fundamental workings in
essentially simple.  Moreover, there are practical consequences of the
simplicity entailed by symmetry:  it is easier to understand the predictions of
physical laws.  For example, working out the details of very-many-body motion
is beyond the reach of actual calculations, even with the help of
computers.  But taking into account the symmetries that are present allows
understanding at least some aspects of the motion, and charting regularities
within it.

Symmetries bring with them conservation laws -- an association that is
precisely formulated by Noether's theorem.  Thus time-translation symmetry,
which states that physical laws do not change as time passes, ensures energy
conservation; space-translation symmetry -- the statement that physical laws
take the same form at different spatial locations -- ensures momentum
conservation.  For another example, we note that quantal description makes use
of complex numbers (involving $\sqrt{-1}$).  But physical quantities are real,
so complex phases can be changed at will, without affecting physical content. 
This invariance against phase redefinition, called {\bf gauge symmetry}, leads
to charge conservation.  The above exemplify a general fact: symmetries
are linked to constants of motion.  Identifying such constants, on the one
hand, satisfies our urge to find regularity and permanence in natural
phenomena, and on the other hand, we are provided with a useful index for
ordering physical data.

However, in spite of our preference that descriptions of Nature be enhanced
by a large amount of symmetry and characterized by many conservation laws, it
must be recognized that actual physical phenomena rarely exhibit overwhelming
regularity.  Therefore, at the very same time that we construct a physical
theory with intrinsic symmetry, we must find a way to break the symmetry in
physical consequences of the model.  Progress in physics can be frequently
seen as the resolution of this tension.

In classical physics, the principal mechanism for symmetry breaking, realized
already within Newtonian mechanics, is through boundary and initial
conditions on dynamical equations of motion -- for example radially
symmetric dynamics for planetary motion allows radially non-symmetric,
non-circular orbits with appropriate initial conditions.  But this mode of
symmetry breaking still permits symmetric configurations - circular orbits,
which are rotationally symmetric, are allowed.  In quantum mechanics, which
anyway does not need initial conditions to make physical predictions, we must
find mechanisms that prohibit symmetric configurations altogether.

In the simplest, most direct approach to symmetry breaking, we suppose that in
fact dynamical laws are not symmetric, but that the asymmetric effects are
``small" and can be ignored ``in first approximation."  The breaking of
rotational symmetry in atoms by an external electromagnetic field or of isospin
symmetry by the small electromagnetic interaction are familiar examples. 
However, this explicit breaking of symmetry is without fundamental interest
for the exact and complete theory; we need more intrinsic mechanisms that work
for theories that actually are symmetric.

A more subtle idea is {\bf spontaneous symmetry breaking}, where the dynamical
laws are symmetric, but only asymmetric configurations are actually realized,
because the symmetric ones are energetically unstable.  This mechanism, urged
upon us by Heisenberg, Anderson, Nambu and Goldstone, is readily illustrated
by the potential energy profile possessing left-right symmetry and depicted in
the Figure.  The left-right symmetric value
at the origin is a point of unstable equilibrium; stable equilibrium is
attained at the reflection unsymmetric points $\pm a$.  Once the system
settles in one or the other location, left-right parity is absent.  One says
that the symmetry of the equations of motion is ``spontaneously'' broken by
the stable solution.

\vskip-.3in
\centerline{\epsffile{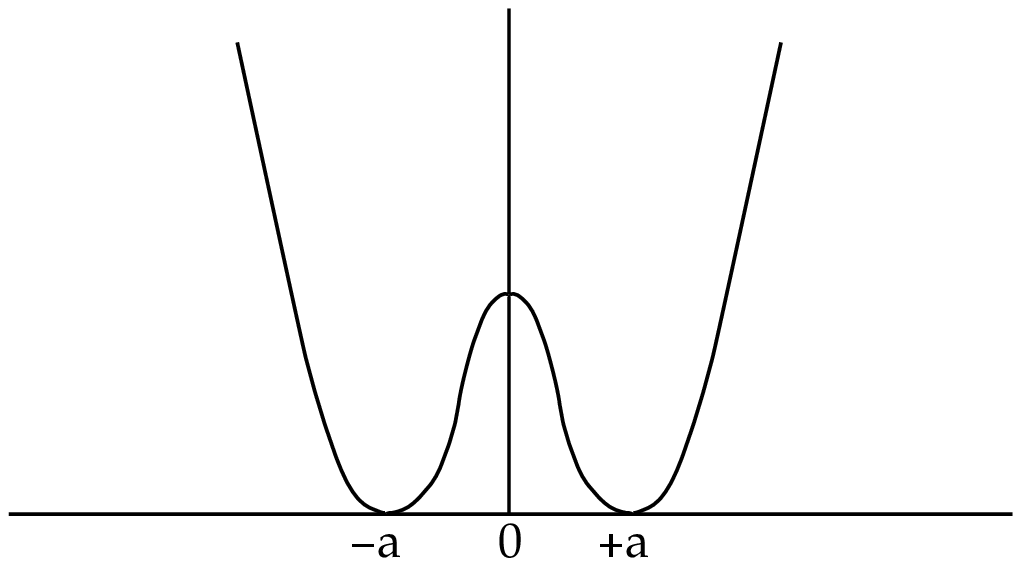}}
\baselineskip=14pt plus .5pt
\noindent{\sf Left-right symmetric particle energy or field theoretic energy
density.  The symmetric point at 0 is energetically unstable.  Stable
configurations are at $\pm a$, and a quantum mechanical particle can tunnel
between them.  In field theory, the energy barrier is infinite, tunneling is
surpressed, the system settles into state $+ a$ or $- a$ and left-right
symmetry is spontaneously broken.}
\newpage

\baselineskip=24pt plus .5pt

But here we come to the first instance where infinities play
a crucial role.  The above discussion of the asymmetric solution is
appropriate to a classical physics description, where a physical state
minimizes energy and is uniquely realized by one or the other configuration at
$\pm a$.  However, quantum mechanically a physical state can comprise a
superposition of classical states, where the necessity of superposing arises
from quantum mechanical tunneling, which allows mixing between classical
configurations.  Therefore if the profile in the Figure describes potential
energy of a single quantum particle as a function of particle position, the 
barrier between the two minima carries finite energy.  The particle can then
tunnel  between the two configurations $\pm a$, and the lowest quantum state
is a superposition, which in the end respects the left-right symmetry. 
Spontaneous symmetry breaking does not occur in quantum particle mechanics.  
However, in a field theory, the graph in the Figure describes spatial energy
{\bf density} as a function of the field, and the total energy barrier is the
finite amount seen in the Figure, multiplied by the infinite spatial volume in
which the field theory is defined.  Therefore the total energy barrier is
infinite, and tunneling is impossible.  Thus spontaneous symmetry breaking can
occur in quantum {\bf field} theory, and Weinberg as well as Salam employed
this mechanism for breaking unwanted symmetries in the standard model.  But we
see that this crucial ingredient of our present-day theory for fundamental
processes is available to us precisely because of the infinite volume of
space, which also is responsible for infrared  divergences!

But infrared problems are not so significant, so let me focus on the
ultraviolet infinities.  These are important for a further, even more subtle
mode of symmetry breaking, which also is crucial for the
phenomenological success of our theories.  This mode of symmetry breaking is
called {\bf anomalous} or {\bf quantum mechanical}, and in order to explain
it, let me begin by reminding that the quantum revolution did not erase our
reliance on the earlier, classical physics.  Indeed, when proposing a theory,
we begin with classical concepts and construct models according to the rules of
classical, pre-quantum physics.  We know, however, such classical reasoning is
not in accord with quantum reality.  Therefore, the classical model is
reanalyzed by the rules of quantum physics, which comprise the true laws of
Nature.  This two-step procedure is called {\bf quantization}.

Differences between the physical pictures drawn by a classical description and
a quantum description are of course profound.  To mention the most dramatic,
we recall that dynamical quantities are described in quantum mechanics by
operators, which need not commute.  Nevertheless, one expects that some
universal concepts transcend the classical/quantal dichotomy, and enjoy
rather the same role in quantum physics as in classical physics.

For a long time it was believed that symmetries and conservation laws of a
theory are not affected by the transition from classical to quantum rules. 
For example if a model possesses translation and gauge invariance on the
classical level, and consequently energy/momentum and charge are conserved
classically, it was believed that after quantization the quantum model is
still translation and gauge invariant so that the energy/momentum and charge
operators are conserved within quantum mechanics, that is, they commute with
the quantum Hamiltonian operator.  But now we know that in general this need
not be so.  Upon quantization, some symmetries of classical physics may
disappear when the quantum theory is properly defined in the presence of its
infinities.  Such tenuous symmetries are said to be {\bf anomalously} broken; 
although present classically, they are absent from the quantum version of the
theory, unless the model is carefully arranged to avoid this effect.

The nomenclature is misleading.  At its discovery, the phenomenon was
unexpected and dubbed ``anomalous."  By now the surprise has worn off, and the
better name today is ``quantum mechanical" symmetry breaking.

Anomalously or quantum mechanically broken symmetries play several and
crucial roles in our present-day physical theories.  In
some instances they save a model from possessing too much symmetry,
which would not be in accord with experiment.  In other instances, the
desire to preserve a symmetry in the quantum theory places strong constraints
on model building and gives experimentally verifiable predictions; more about
this later. [\ref{ref:3}]

Now I shall describe two specific examples of the anomaly phenomenon. 
Consider first massless fermions moving in an electromagnetic field
background.  Massive, spin-$1\over 2$ fermions possess two spin states -- up
and down -- but massless fermions can exist with only one spin state, called a
{\bf helicity} state, in which spin is projected along the direction of
motion.  So the massless fermions with which we are here concerned carry only
one helicity and these are an ingredient in present-day theories of quarks and
leptons.  Moreover, they also arise in condensed matter physics, not because
one is dealing with massless, single-helicity particles, but because a
well-formulated approximation to various many-body Hamiltonians can result in
a first order matrix equation that is identical to the equation for
single-helicity massless fermions, {\it i.e.}, a massless Dirac-Weyl equation
for a spinor $\Psi$.

If we view the spinor field $\Psi$ as an ordinary mathematical function, we
recognize that it possesses a complex phase, which can be altered without
changing the physical content of the equation that $\Psi$ obeys.  We expect
therefore that this instance of gauge invariance implies charge conservation. 
However, in a quantum field theory $\Psi$ is a quantized field operator, and
one finds that in fact the charge operator $Q$ is not conserved; rather

\beq
{dQ \over dt} = {i \over \hbar} [H,Q] \propto \int_{\rm volume}
\hbox{\bf E} \cdot \hbox{\bf B}
\label{eq:1}
\eeq    
where {\bf E} and {\bf B} are the background electric and magnetic fields in
which our massless fermion is moving; gauge invariance is lost! 

One way to understand this breaking of symmetry is to observe that our model
deals with {\bf massless} fermions and conservation of charge for {\bf
single-helicity\/} fermions makes sense only if there are no fermion masses. 
But quantum field theory is beset by its ultraviolet infinities that must be
controlled in order to do a computation.  This is accomplished by
regularization and renormalization, which introduces mass scales for the
fermions, and we see that the symmetry is anomalously broken by the
ultraviolet infinities of the theory.

The phase-invariance of single-helicity fermions is called {\bf chiral
symmetry} and chiral symmetry has many important roles in the standard model,
which involves many kinds of fermion fields, corresponding to the various
quarks and leptons.  In those channels where a gauge vector meson couples to
the fermions, chiral symmetry must be maintained to ensure gauge invariance. 
Consequently fermion content must be carefully adjusted so that the anomaly
disappears.  This is achieved, because the proportionality constant in
the failed conservation law (\ref{eq:1}) involves a sum over all the fermion
charges, $\sum\limits_n q_n$, so if that quantity vanishes the anomaly is
absent.  In the standard model the sum indeed vanishes, separately for each of
the three fermion families.  For a single family this works out as follows:

\begin{center}
\begin{tabular}{llcrlr}
three quarks & $\qquad  q_n$   &= & ${2 \over 3}$ & $\Rightarrow$ & $2$ \\
three quarks &  $\qquad  q_n$  &= & $-{1 \over 3}$ & $ \Rightarrow$ & $ -1$ \\
one~lepton & $\qquad  q_n$ &= & $-1$ & $ \Rightarrow $ & $-1$ \\
one~lepton & $\qquad  q_n$ &= & $0 $ & $\Rightarrow$ & $ 0$ \\
              &            &  & $\sum\limits_n q_n$ &  =   & $ 0$
\end{tabular}
\end{center}

In channels to which no gauge vector meson couples, there is no requirement
that the anomaly vanish, and this is fortunate: a theoretical analysis
shows that gauge invariance in the up-down quark channel prohibits the
two-photon decay of the neutral pion (which is composed of up and down
quarks).  But the decay does occur with the
invariant decay amplitude of $(.025 \pm .001)(GeV)^{-1}$. Before anomalous
symmetry breaking was understood, this decay could not be fitted into the
standard model, which seemed to possess the decay-forbidding
chiral symmetry.  Once it was realized that the relevant chiral symmetry is
anomalously broken, this obstacle to phenomenological viability of the standard
model was removed.  Indeed since the anomaly is completely known, the decay
amplitude can be completely calculated (in the approximation that the pion is
massless) and one finds
$(.025)(GeV)^{-1}$, in excellent agreement with experiment.  

We must conclude that Nature knows about and makes use of the anomaly
mechanism: fermions are arranged into gauge-anomaly-free representations,
and the requirement that anomalies disappear ``explains" the charges of
elementary fermions; the pion decays into two photons because of an anomaly in
an ungauged channel.  It seems therefore that in local quantum field theory
these phenomenologically desirable results are facilitated by ultraviolet
divergences, which give rise to symmetry anomalies.

The observation that infinities of quantum field theory lead to anomalous
symmetry breaking allows comprehending a second
example of quantum mechanical breaking of yet another symmetry -- that of scale
invariance.  Like the space-time translations mentioned earlier, which lead to
energy-momentum conservation, scale transformations also act on space-time
coordinates, but in a different manner: they dilate the coordinates, thereby
changing the units of space and time measurements.  Such transformations will
be symmetry operations in models that possess no fundamental parameters with
time or space dimensionality, and therefore do not contain an absolute scale
for units of space and time.  Our quantum chromodynamical model (QCD) for
quarks is free of such dimensional parameters, and it would appear that this
theory is scale invariant -- but Nature certainly is not!  The observed
variety of different objects with different sizes and masses exhibits many
different and inequivalent scales.  Thus if scale symmetry of the {\it
classical} field theory, which underlies the {\it quantum} field theory of
QCD, were to survive quantization, experiment would have grossly contradicted
the model, which therefore would have to be rejected.  Fortunately, scale
symmetry is quantum mechanically broken, owing to the scales that are
introduced in the regularization and renormalization of ultraviolet
singularities.  Once again a quantum field theoretic pathology has a physical
effect, a beneficial one: an unwanted symmetry is anomalously broken, and
removed from the theory.

Another application of anomalously broken scale invariance, especially as
realized in the renormalization group program, concerns high energy behavior
in particle physics and critical phenomena in condensed matter physics.  A
scale-invariant quantum theory could not describe the rich variety of
observed effects, so it is fortunate that the symmetry is quantum
mechanically broken. [\ref{ref:4}]

A different perspective on the anomaly phenomenon comes from
the path integral formulation of quantum theory, where one integrates over
classical paths the phase exponential of the classical action.

\beq
{\rm Quantum~Mechanics} \Longleftrightarrow 
\int_{\rm\textstyle (measure~on~paths)} e^{i/\hbar {\rm\textstyle
(classical~action)}}
\label{eq:2}
\eeq
When the classical action possess a symmetry, the quantum theory will respect
that symmetry if the measure on paths is unchanged by the relevant
transformation.  In the known examples (chiral symmetry, scale symmetry)
anomalies arise precisely because the measure fails to be invariant and this
failure is once again related to infinities: the measure is an {\bf infinite}
product of measure elements for each point in the space-time where the quantum
(field) theory is defined; regulating this infinite product destroys its
apparent invariance.

Yet another approach to chiral anomalies, which arise in (massless)
fermion theories, makes reference to the first instance of
regularization/renormalization, used by Dirac to remove the negative-energy
solutions to his equation.  Recall that to define a quantum {\bf field}
theory of fermions, it is necessary to fill the negative-energy sea and to
renormalize the infinite mass and charge of the filled states to zero.  In
modern formulations this is achieved by ``normal ordering" but for our
purposes it is better to remain with the more explicit procedure of
subtracting the infinities, {\it i.e.} renormalizing them.

It can then be shown that in the presence of a gauge field, the distinction
between ``empty" positive-energy states and ``filled" negative-energy states
can not be drawn in a gauge invariant manner, for massless, single-helicity
fermions.  Within this framework, the chiral anomaly comes from the gauge
non-invariance of the infinite negative-energy sea.  Since anomalies have
physical consequences, we must assign physical reality to this infinite
negative-energy sea. [\ref{ref:5}]

Actually, in condensed matter physics, where a Dirac-type equation governs
electrons, owing to a linearization of dynamical equations near the Fermi
surface, the negative-energy states {\bf do} have physical reality: they
correspond to filled, bound states, while the positive energy states
describe electrons in the conduction band.  Consequently, chiral anomalies
also have a role in condensed matter physics, when the system is idealized so
that the negative-energy sea is taken to be infinite. [\ref{ref:6}] 
 
In this condensed matter context another curious, physically realized,
phenomenon has been identified.  When the charge of the filled negative states
is renormalized to zero, one is subtracting an infinite quantity and rules
have to be agreed upon so that no ambiguities arise when infinite quantities
are manipulated.  With this agreed-upon subtraction procedure, the charge of
the vacuum is zero, and filled states of positive energy carry integer units
of charge.  Into the system one can insert a soliton -- a localized structure
that distinguishes between different domains of the condensed matter.  In the
presence of such a soliton, one needs to recalculate charges using the
agreed-upon rules for handling infinities and one finds, surprisingly, a
non-integer result, typically half-integer: the negative-energy sea is
distorted by the soliton to yield a half a unit of charge.  The existence
of fractionally charged states in the presence of solitons has been
experimentally identified in polyacetylene.~[\ref{ref:7}]  We thus have another
example of a physical effect emerging from infinities of quantum field theory.

Let me conclude my qualitative discussion of anomalies with an
explicit example from quantum mechanics, whose wave functions provide a link
between particle and field theoretic dynamics.  My example also dispels any
suspicion that ultraviolet divergences and the consequent anomalies are tied to
the complexities of relativistic quantum field theory: the non-relativistic
example shows that locality is what matters.

Recall first the basic dynamical equation of quantum mechanics: the
time independent Schr\"odinger equation for a particle of mass $m$ moving
in a potential $V$({\bf r}) with energy $E$.
\beq
\left( - \nabla^2 + {2m \over \hbar^2} V (\hbox{\bf r}) \right) \psi
(\hbox{\bf r}) =  {2m \over \hbar^2} E \psi (\hbox{\bf r})  \,\, .
\label{eq:3}
\eeq
In its most important physical applications, this equation is taken in three
spatial dimensions and $V (\hbox{\bf r})$ is proportional to $1/r$ for the
Coulomb force relevant in atoms.  Here we want to take a different model with
potential that is proportional to the inverse square, so that the
Schr\"odinger equation is presented as
\beq
\left( - \nabla^2 + {\lambda \over r^2} \right) \psi(\hbox{\bf r})  = 
k^2  \psi (\hbox{\bf r})  \,\, , \qquad 
k^2 \equiv {2m \over \hbar^2}E \,\, .
\label{eq:4}
\eeq
In this model, transforming the length scale is a symmetry: because the
Laplacian scales as $r^{-2}$, $\lambda$ is dimensionless and in
(\ref{eq:4}) there is no intrinsic unit of length.  A consequence of scale
invariance is that the scattering phase shifts and the $S$ matrix, which in
general depend on energy, {\it i.e.} on $k$, are energy-independent in scale
invariant models, and indeed when the above Schr\"odinger equation is solved,
one verifies this prediction of the symmetry by finding an energy-independent
$S$ matrix.  Thus scale invariance is maintained in this example -- there are
no surprises.

Let us now look to a similar model, but in two dimensions with a
$\delta$-function potential, which localizes the interaction at a point.
\beq
\Bigg( - \nabla^2 + \lambda \delta^2 (\hbox{\bf r}) \Bigg) \psi
(\hbox{\bf r}) =  k^2\psi (\hbox{\bf r})  \,\, .
\label{eq:5}
\eeq
Since in two dimensions the two-dimensional $\delta$-function scales as
$1/r^2$, the above model also appears scale invariant; $\lambda$ is
dimensionless.  But in spite of the simplicity of the local contact
interaction, the Schr\"odinger equation suffers a short-distance, ultraviolet
singularity at {\bf r}=0, which must be renormalized.  Here is not the place
for a detailed analysis, but the result is that only the $s$-wave possesses a
non-vanishing phase shift $\delta_0$, which shows a logarithmic dependence on
energy.
\beq
{\rm ctn}~\delta_0 = {2 \over \pi} \ln kR + {1 \over \lambda}
\label{eq:6}
\eeq
$R$ is a scale that arises in the renormalization, and scale symmetry is
decisively and quantum mechanically broken. [\ref{ref:8}]  Moreover, the
scattering is non-trivial solely as a consequence of broken scale invariance.
It is easily verified that the two-dimensional $\delta$-function in classical
theory, where it {\bf is} scale invariant, produces no scattering.

Furthermore the $\delta$-function model may be second quantized, by promoting
the wave-function to a field operator $\hat{\psi}$, and positing a field
theoretic Hamiltonian density operator of the form
\beq
{\cal H} = {\hbar^2 \over 2m} \hbox{\boldmath $\nabla$} \hat{\psi}^\ast \cdot
\hbox{\boldmath $\nabla$}
\hat{\psi} + {\lambda \over 2} (\hat{\psi}^\ast \hat{\psi})^2
\label{eq:7}
\eeq
The physics of a second-quantized non-relativistic field theory is the same
as that of many body quantum particle mechanics, and the two-body problem (with
center-of-mass motion removed) is governed by the above-mentioned
Schr\"odinger equation: the $\delta$-function interaction is encoded in the
${\lambda \over 2} (\hat{\psi}^\ast \hat{\psi})^2$ interaction term, which is
local.

Analysis of the field theory confirms its apparent scale invariance,
but the local interaction produces ultraviolet divergences, that must be
regulated and renormalized, thereby effecting an anomalous, quantum
mechanical breaking of the scale symmetry. [\ref{ref:9}]

The above list of examples persuades me that the infinities of local quantum
field theory -- be they ultraviolet divergences, or an infinite functional
measure, or the infinite negative-energy sea -- are not merely
undesirable blemishes on the theory, which should be hidden or --
even better -- not present in the first place.  On the contrary the 
successes of various field theories in describing physical phenomena depend on
the occurrence of these infinities.  One cannot escape the conclusion that
Nature makes use of anomalous symmetry breaking, which occurs in local field
theory owing to underlying infinities in the mathematical description.

The title of my talk expresses the surprise at this state of affairs: surely
it is unreasonable that some of the effectiveness of quantum field theory
derives from its infinities.  Of course my title is also a riff on Wigner's
well-known aphorism about the unreasonable effectiveness of mathematics in
physics.[\ref{ref:10}]  We can understand the effectiveness of mathematics: it
is the language of physics and {\bf any} language is effective in expressing
the ideas of its subject.  Field theory, in my opinion, is also a language, a
more specialized language that we have invented for describing fundamental
systems with many degrees of freedom.  It follows that all relevant phenomena
will be expressible in the chosen language, but it may be that some features
can be expressed only awkwardly.  Thus for me chiral and scale symmetry
breaking are completely natural effects, but their description in our present
language -- quantum field theory -- is awkward and leads us to extreme
formulations,  which make use of infinities.  One hopes that there is a more
felicitous description, in an as yet undiscovered language.  It is striking
that anomalies afflict precisely those symmetries that depend on absence of
mass: chiral symmetry, scale symmetry.  Perhaps when we have a natural
language for anomalous symmetry breaking we shall also be able to speak in a
comprehensible way about mass, which today remains a mystery. 

Some evidence for a description of anomalies without the paradox of field
theoretic infinities comes from the fact that they have a very natural
mathematical expression.  For example the ${\bf E} \cdot {\bf B}$ in our
anomalous non-conservation of chiral charge is an example of a
Chern-Pontryagin density, whose integral measures the topological index of
gauge field configurations and enters in the Atiyah-Singer index theorem. 
Also the fractional charge phenomenon, which physicists found in Dirac's
infinite negative energy sea, can alternatitively be related to the
mathematicians' Atiyah-Patodi-Singer spectral flow.[\ref{ref:11}]

The relation between mathematical entities and field theoretical anomalies
was realized twenty years ago and has led to a flourishing interaction between
physics and mathematics, which today culminates in the string program. 
However, it seems to me that here mathematical ideas have taken the lead,
without advancing our physical understanding -- the string program thus far
has not illuminated physical questions.  In particular I wonder where within
the completely finite and non-local dynamics of string theory are we to find
the mechanisms for symmetry breaking that are needed in order to explain the
world around us. 

\newpage
\section*{NOTES}
\frenchspacing

\baselineskip=14pt plus .5pt

\begin{enumerate}

\item \label{ref:1} For an account of the origins of classical field theory,
see L.P.~Williams, {\it The Origins of Field Theory\/} (Random House, New
York, NY 1966).

\item \label{ref:2} For an account of the origins of quantum field theory,
specifically quantum electrodynamcis, see S.S.~Schweber, {\it QED and the Men
Who Made It\/} (Princeton University Press, Princeton, NJ 1994).

\item \label{ref:3} For an account of quantum anomalies in conservation laws,
see R.~Jackiw ``Field Theoretic Investigations in Current Algebra" and
``Topological Investigation of Quantized Gauge Theories" in S.~Treiman,
R.~Jackiw, B.~Zumino and E.~Witten, {\it Current Algebra and Anomalies\/}
(Princeton University Press/World Scientific, Princeton, NJ/Singabpore 1985).

\item \label{ref:4} For a review, see K.~Wilson, ``The Renormalization Group,
Critical Phenomena and the Kondo Problem," {\it Rev.~Mod.~Phys.\/} {\bf 47},
773 (1975).

\item \label{ref:5}
 See R.~Jackiw, ``Effect of Dirac's Negative Energy Sea on Quantum Numbers,''
{\it Helv.~Phys.~Acta\/} {\bf 59}, 835 (1986).

\item \label{ref:6}
For a selection of applications, see H.B.~Nielsen and M.~Ninomiya, ``The
Adler-Bell-Jackiw Anomaly and Weyl Fermions in a Crystal," {\it
Phys.~Lett.\/}~{\bf 130B}, 389 (1983); I.~Krive and A.~Rozhavsky, ``Evidence
for a Chiral Anomaly in Solid State Physics," {\it Phys.~Lett.\/}~{\bf 113A},
313 (1983); M.~Stone and F.~Gaitan, ``Topological Charge and Chiral Anomalies
in Fermi Superfluids,'' {\it Ann.~Phys.\/} (NY) {\bf 178}, 89 (1987). 

\item \label{ref:7} See R.~Jackiw and J.R.~Schrieffer,``Solitons with Fermion
Number $1/2$ in Condensed Matter and Relativistic Field Theories,'' {\it
Nucl.~Phys.\/} {\bf B190} [FS3], 253 (1981) and R.~Jackiw, ``Fractional
Fermions," {\it Comments Nucl.~Part.~Phys.\/} {\bf 13}, 15 (1984).

\item \label{ref:8} See R.~Jackiw, ``Introducing Scale Symmetry'' {\it
Physics Today} {\bf 25}, No.~1, 23 (1972) and ``Delta-Function Potentials in
Two- and Three- Dimensional Quantum Mechanics'' in {\it M.A.B.~B\'eg
Memorial Volume\/}, A.~Ali and P.~Hoodbhoy eds.~(World Scientific, Singapore
1991).

\item \label{ref:9} O.~Bergman, ``Non-Relativistic Field Theoretic Scale
Anomaly,'' {\it Phys.~Rev.~D\/} {\bf 46}, 5474 (1992); O.~Bergman and
G.~Lozano, ``Aharonov-Bohm Scattering, Contact Interactions and Scale
Invariance,'' {\it Ann.~Phys.\/} (NY) {\bf 229}, 416 (1994); B.~Holstein,
``Anomalies for Pedestrians.'' {\it Am.~Jnl.~Phys.\/} {\bf 61}, 142 (1993).

\item \label{ref:10} E.P.~Wigner, ``The Unreasonable Effectiveness of
Mathematics in the Natural Sciences;" see also his ``Symmetry and
Conservation Laws'' and ``Invariance in Physical Theory.''  All three articles
are reprinted in {\it The Collected Works of Eugene Paul Wigner}, Vol. VI, Part
B,  J. Mehra ed. (Springer Verlag, Berlin, 1995).

\item \label{ref:11} For a review, see A.J. Niemi and G. Semenoff, ``Fermion
Number Fractionization in Quantum Field Theory,'' {\it Physics Rep.\/} {\bf
135}, 99~(1986).

\end{enumerate}
\nonfrenchspacing

\end{document}